\documentclass[twocolumn,preprintnumbers,amsmath,amssymb,showpacs]{revtex4}
\usepackage{graphicx}
\usepackage{dcolumn}
\usepackage{bm}

\newcommand{\remark}[1]{{\tt[#1]}}

\newcommand{\whatref}[1]{ \remark{\mbox{$\backslash$ref\{?\}}}}

\begin{document}

\title{Stabilization of nonlinear velocity profiles in 
athermal systems undergoing planar shear flow}

\author{Ning Xu$^1$}
\author{Corey S. O'Hern$^{1,2}$}
\author{Lou Kondic$^3$}
\address{$^1$~Department of Mechanical Engineering, 
Yale University, New Haven, CT 06520-8284.\\
$^2$~Department of Physics, Yale University, New Haven, CT 06520-8120.\\
$^3$~Department of Mathematical Sciences, New Jersey Institute 
of Technology, Newark, NJ  07102.}
\date{\today}

\begin{abstract}
We perform molecular dynamics simulations of model granular systems
undergoing boundary-driven planar shear flow in two spatial dimensions
with the goal of developing a more complete understanding of how dense
particulate systems respond to applied shear.  In particular, we are
interested in determining when these systems will possess linear
velocity profiles and when they will develop highly localized velocity
profiles in response to shear.  In previous work on similar
systems we showed that nonlinear velocity profiles form when the
speed of the shearing boundary exceeds the speed of shear waves in the
material.  However, we find that nonlinear velocity profiles in these
systems are unstable at very long times.  The degree of nonlinearity
slowly decreases in time; the velocity profiles become linear when the
granular temperature and density profiles are uniform across the
system at long times.  We measure the time $t_l$ required for the
velocity profiles to become linear and find that $t_l$ increases as a
power-law with the speed of the shearing boundary and increases
rapidly as the packing fraction approaches random close packing.  We
also performed simulations in which differences in the granular
temperature across the system were maintained by vertically vibrating
one of the boundaries during shear flow.  We find that nonlinear
velocity profiles form and are stable at long times if the difference
in the granular temperature across the system exceeds a threshold
value that is comparable to the glass transition temperature in an
equilibrium system at the same average density.  Finally, the sheared
and vibrated systems form stable shear bands, or highly localized
velocity profiles, when the applied shear stress is lowered below the
yield stress of the static part of the system.
\end{abstract}

\maketitle

\section{Introduction}

The response of fluids to applied shear is an important and well
studied problem.  When a Newtonian fluid is slowly sheared, for
example by moving the top boundary of the system at fixed velocity $u$
in the $x$-direction at height $y=L_y$ relative to a stationary bottom
boundary at $y=0$, a linear velocity profile $v_x(y) = {\dot \gamma}
y$ is established, where the shear rate is ${\dot \gamma} = u/L_y$.
In addition, in the small shear rate limit the shear stress is
linearly related to the shear rate $\sigma_{xy} = \eta {\dot \gamma}$,
where $\eta$ is the shear viscosity.  This relation is often employed
to measure the shear viscosity of simple fluids.

However, what is the response of non-Newtonian fluids like granular
materials to an applied shear?  In contrast to simple liquids, it is
extremely difficult to predict the response of dense granular media to
shear because they are inherently out of thermal equilibrium, interact
via frictional and enduring contacts, and possess a nonzero yield
stress.  Granular materials do not flow homogeneously when they are
sheared, instead, shear is often localized into shear bands.  When
this occurs, most of the flow is confined to a narrow, locally dilated
region near the shearing boundary while the remainder of the system is
nearly static.  Recent experimental work investigating the response of
dense granular materials to shear includes studies of Couette flow in
2D \cite{howell}, in 3D for spherical particles
\cite{losert,mueth2} and as a function of particle shape \cite{mueth},
studies of cyclic planar shear \cite{mueggenburg}, studies of wide
shear zones formed in the bulk using a modified Couette geometry
\cite{fenistein}, and studies of chute flow \cite{pouliquen}.

Recent theoretical studies \cite{jenkins} have shown that kinetic
theory can correctly predict the velocity, granular temperature, and
density profiles found in experiments of Couette flow in the dilute
regime \cite{wildman}.  However, kinetic theory and modifications
included to account for diverging viscosity are not analytically
tractable, and are unlikely to predict accurately the properties of
dense shear flows.  Thus, molecular dynamics simulations are often
employed to study dense granular shear flows, for example in
Refs.~\cite{thompson,aharonov,volfson,dacruz}.  We choose a similar
plan of attack and perform molecular dynamics
simulations of model frictionless dense granular systems undergoing
boundary-driven planar shear flow in 2D.  We are interested in
answering several important questions: What are the velocity, granular
temperature, and density profiles as a function of the velocity of the
shearing boundary?  In particular, do highly localized velocity
profiles form and, if so, are they stable at long times?  We will
investigate these questions in simple 2D systems composed of inelastic
but frictionless particles in planar shear cells, and in the absence of
gravity.  Our intent is to understand in detail the time and spatial
dependence of the response in these simple systems to shear first, and
then extend our studies to 3D systems, systems composed of frictional
particles, and Couette shear cells.  In Sec.~\ref{sec:future}, we
present preliminary results from these future studies.

We previously reported that nonlinear velocity profiles form in
repulsive athermal systems when the velocity of the shearing wall
exceeds the speed of shear waves in the material \cite{xu}.  In the
present article, we study much longer time scales and show that
nonlinear velocity profiles are unstable at long times.  In addition,
the granular temperature and density profiles become uniform
throughout the system at long times.  Thus, granular temperature and
density gradients cannot be maintained by planar shear flow in dense
systems with dissipative but frictionless interactions. We measure the
time $t_l$ for the velocity profiles to become linear and find that
$t_l$ scales as a power-law in the velocity of the shearing boundary,
and increases rapidly as the average density of the system approaches
random close packing $\phi_{\rm rcp}$ from above, where $\phi_{\rm
rcp} \approx 0.84$ in 2D \cite{ohern}.

To maintain a granular temperature difference across the system, we also
studied systems that were both sheared and vertically vibrated.  We find
that if the difference in the granular temperature across the system
exceeds a threshold that is comparable to the glass transition
temperature in an equilibrium system at the same average density,
nonlinear velocity profiles are stable at long times.  Finally, we
show that shear bands, or highly localized velocity profiles, form in
the vibrated and sheared systems if the shear stress is tuned below
the yield stress of the static part of the system.

\section{Methods}     
\label{sec:methods}

Before discussing our results further, we will first describe our
numerical model and methods.  We performed molecular dynamics
simulations of purely repulsive and frictionless athermal systems
undergoing boundary-driven planar shear flow in two spatial
dimensions.  The systems were composed of $N/2$ large and $N/2$ small
particles with equal mass $m$ and diameter ratio $1.4$ to prevent
crystallization and segregation during shear.  The starting
configurations were prepared by choosing an average packing fraction
and random initial positions and then allowing the system to relax to
the nearest local potential energy minimum \cite{ohern} using the
conjugate gradient method \cite{numrec}.  During the quench,
periodic boundary conditions were implemented in both the $x$- and
$y$-directions.  Following the quench, particles with $y$-coordinates
$y>L_y$ ($y<0$) were chosen to comprise the top (bottom) boundary.
Thus, the top and bottom walls were rough and amorphous.

Shear flow in the $x$-direction with a shear gradient in the
$y$-direction and global shear rate $u/L_y$ was created by moving all
particles in the top wall at fixed velocity $u$ in the $x$-direction
relative to the stationary bottom wall.  During shear flow, periodic
boundary conditions were imposed in the $x$-direction.  We chose an
aspect ratio $L_x/L_y=1/4$ with more than $50$ particles along the
shear gradient direction to reduce finite-size effects, and focused on
systems with packing fractions in the range $\phi=[0.835,0.95]$.

Both bulk particles and particles comprising the boundary interact via
the purely repulsive harmonic spring potential
\begin{equation}
V(r_{ij}) =\frac{\epsilon}{2} \left(1-\frac{r_{ij}}{\sigma_{ij}} \right)^2
\Theta \left( \frac{\sigma_{ij}}{r_{ij}}-1 \right),
\label{equ:potential}
\end{equation}
where $\epsilon$ is the characteristic energy scale of the
interaction, $\sigma_{ij}=(\sigma_i + \sigma_j)/2$ is the average
diameter of particles $i$ and $j$, $r_{ij}$ is their separation, and
$\Theta(x)$ is the Heaviside step function.  Note that the interaction
potential is zero when $r_{ij} \ge \sigma_{ij}$.

In our simulations, we employ athermal or dissipative dynamics with no
frictional or tangential forces \cite{herrmann}.  The position and
velocity of particles in the bulk were obtained by solving
\begin{equation} 
\label{dissipative} 
m \frac{d^2{\vec r}_i}{dt^2} = {\vec F}^r_i - b_n \sum_j \left[ 
\left({\vec v}_{i} 
- {\vec v}_{j}\right) \cdot \hat{r}_{ij} \right] \hat{r}_{ij}, 
\end{equation} 
where ${\vec F}^r_i=-\sum_j dV(r_{ij})/dr_{ij} {\hat r}_{ij}$, the
sums over $j$ only include particles that overlap $i$, ${\vec v}_{i}$
is the velocity of particle $i$, and $b_n>0$ is the damping
coefficient.  We focus on underdamped systems in this study;
specifically, for most simulations we use $b_n=0.0375$ (coefficient of
restitution $e=0.92$).  In addition, we show some results that cover a
range of $e=[0.1,0.99]$.  The units of length, energy, and time are
chosen as the small particle diameter $\sigma$, $\epsilon$, and
$1/\omega_c \equiv \sigma \sqrt{m/\epsilon}$, and all quantities were
normalized by these.

We also performed simulations in which the system was both sheared and
vertically vibrated.  As discussed above, the system was sheared by
constraining the top boundary to move in the $x$-direction at fixed
speed $u$ relative to the bottom boundary.  In addition, particles in
either the top or bottom boundary were vibrated vertically such that
the $y$-coordinates of the boundary particles varied in time as
\begin{equation}
\label{vibration}
y_i = y_{i0} + A \sin \left( \omega t \right),
\end{equation}   
where $y_{i0}$ is the initial position of boundary particle $i$,
$A$ is the amplitude, and $\omega$ is the angular frequency of the
vibration.  We fixed $A=\sigma/5$ so that the vibrations do not lead
to large, unphysical particle overlaps and $\omega$ was tuned over a
range of frequencies.  Note that in our choice of units $\omega$ is
normalized by the natural frequency $\omega_c$ of the linear spring
interactions.  It should also be pointed out that the simulations with
both shear and vibration are not performed at constant volume but at
fixed average volume.

We measured several physical quantities in these simulations including
the local flow velocity $v_x$, packing fraction $\phi$, and velocity
fluctuations or granular temperature $(\delta v_y)^2 = \langle v_y^2
\rangle - \langle v_y \rangle^2$ in the shear gradient direction, as a
function of the boundary velocity $u$ and vibration frequency
$\omega$.  To study properties as a function of the vertical distance
from the fixed boundaries, we divided the system into rectangular bins
centered at height $y$ and averaged the quantities over the height of
the bin $\Delta y \approx 2$ large particle diameters.  To improve the
statistics and to study time dependence, we also
performed ensemble averages over at least $50$ different initial
configurations.  In the discussion below, we will denote ensemble
averages using $\langle . \rangle$.  When the systems are stationary
but still fluctuate in time, we also perform time averages and these
are denoted by $\langle . \rangle_t$.

\begin{figure}
\scalebox{0.5}{\includegraphics{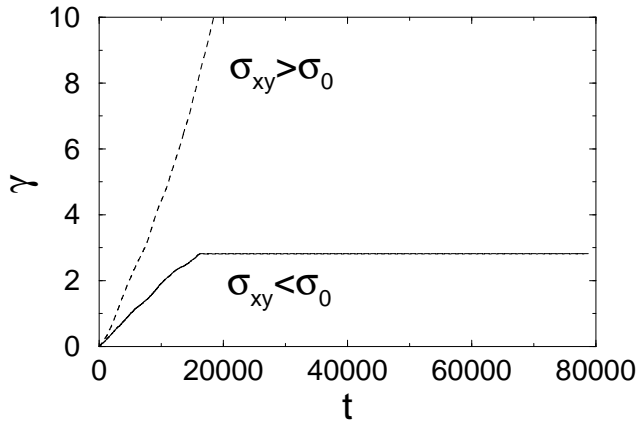}}%
\vspace{-0.15in}
\caption{\label{fig:yield} Shear strain $\gamma = x/L_y$ versus time
$t$ for two shear stresses $\sigma_{xy} = 1.57 \times 10^{-3}$ (solid line)
and $1.80 \times 10^{-3}$ (dashed line) applied to a system at $\phi=0.90$
that was unsheared at $t=0$.}
\vspace{-0.1in} 
\end{figure}
  
We also monitored the shear stress $\sigma_{xy}$ during the course of
our simulations.  The total shear stress in the bulk can be calculated
using the virial expression
\begin{equation}
\label{stress}
\sigma_{xy} = - \frac{1}{L_xL_y}\left( \sum_i \delta v_{i,x} \delta
v_{i,y} + \sum_{i>j} r_{ij,x} F_{ij,y} \right),
\end{equation}
where $r_{ij,x}$ is the $x$-component of ${\vec r}_{ij} = {\vec r}_i -
{\vec r}_j$, $F_{ij,y}$ is the $y$-component of the total pair force
${\vec F}_{ij}$ including both conservative and damping forces, and
$\delta v_{i,x}$ and $\delta v_{i,y}$ measure deviations in the
velocity of a particle from the velocity averaged over the entire
system.  We also compared the shear stress obtained from
Eq.~\ref{stress} with the shear stress $|F^B_x|/L_x$ on the
boundaries, where $F^B_x$ is the total force in the $x$-direction
acting on either the top or bottom boundary.  On average, the bulk
shear stress in Eq.~\ref{stress} and the shear stress on the
boundaries were within a few percent of each other.

\begin{figure}
\scalebox{0.5}{\includegraphics{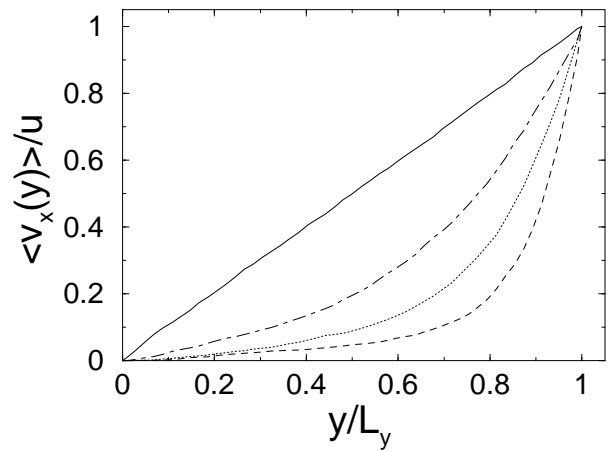}}%
\vspace{-0.15in}
\caption{\label{fig:vp_strain} Ensemble averaged velocity $\langle
v_x(y) \rangle$ in the shear flow direction as a function of height
$y/L_y$ from the stationary boundary for a system sheared at $u=0.364$
with average packing fraction $\phi=0.90$ and $b_n = 0.0375$ at
several different times $t > t_s \approx 470$, where $t_s$ is the time
required for a shear wave to traverse the system.  The curves
correspond to $t=520$ (dashed line), $1000$ (dotted line), $2000$
(dot-dashed line), and $20000$ (solid line).  These times correspond
to large shear strains $\gamma = 2.6$, $5$, $10$, and $100$. Each
curve was averaged over an ensemble of at least $50$ different
starting configurations.}
\vspace{-0.1in} 
\end{figure}

To make contact with the recent results on sheared Lennard-Jones
glasses in Ref.~\cite{varnik}, we measured the yield stress of the
static starting configurations.  To do this, we applied a constant
horizontal force $F$ (or shear stress $\sigma_{xy} = F/L_x$) to the
top boundary.  Results from the simulations at constant horizontal
force are shown in Fig.~\ref{fig:yield}.  Initially, the system flows.
However, when the shear stress is below the yield stress, the system
finds a state that can sustain the applied shear stress and stops
flowing.  If the applied shear stress is above the yield stress, the
system will flow indefinitely. We defined the yield shear stress
$\sigma_0$ as the minimum shear stress above which the shear strain
continues to increase beyond $\gamma = 10$.

\section{Results}

In this section, we report the results from two sets of numerical
simulations of purely repulsive and frictionless athermal systems.
Section~\ref{sec:unstable} presents results from simulations of
boundary-driven planar shear flow.  We show that nonlinear velocity
profiles form at short time scales, but they slowly evolve into linear
profiles at long times.  As the velocity profiles evolve toward linear
ones, the local granular temperature and packing fraction become
uniform.  Section~\ref{sec:vibration} presents results from
simulations of boundary-driven planar shear flow in the presence of
vertical vibrations.  We find that highly nonlinear velocity profiles
can be stabilized at long times if a sufficiently large granular
temperature difference is maintained across the system.

\begin{figure}
\scalebox{0.5}{\includegraphics{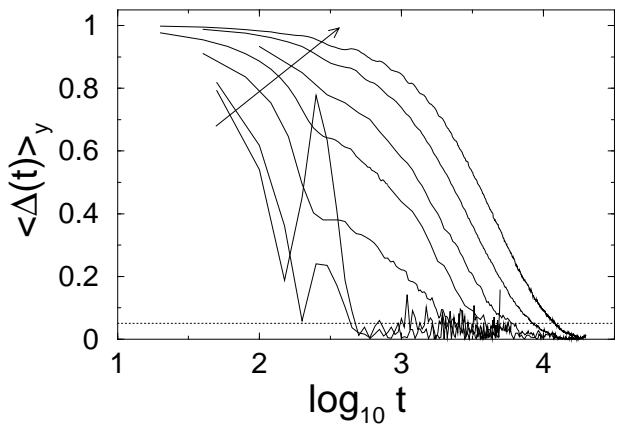}}%
\vspace{-0.15in}
\caption{\label{fig:vevol} Degree of nonlinearity $\langle
\Delta(t) \rangle_y$ of the velocity profile as a function of time $t$
at several different boundary velocities in a system at $\phi = 0.90$
and $b_n = 0.0375$.  We show boundary velocities $u=0.00727$,
$0.0145$, $0.0364$, $0.0727$, $0.145$, $0.364$, $u=0.727$ over two
orders of magnitude; $u$ increases from left to right as shown by the
arrow.  The dotted line $\langle \Delta(t) \rangle_y=0.05$ was used to
estimate the time $t_l$ required for the system to attain an
approximately linear velocity profile.  The spike in $\langle
\Delta(t) \rangle_y$ at small $u$ and short times occurs because the
velocity profile can fluctuate above the linear profile.}
\vspace{-0.1in} 
\end{figure}

\subsection{Time evolution of velocity profiles}
\label{sec:unstable}

In our previous studies of boundary-driven planar shear flow
\cite{xu}, we reported that nonlinear velocity profiles form 
when the velocity $u$ of the shearing boundary exceeds $u_c =
u_s/2$, where $u_s$ is speed of shear waves in the system.  This
condition was obtained by comparing the time $t_s = 2 L_y/u_s$ for a
shear wave to traverse the system to the time $t_u = L_y/u$ for the
system to shear unit strain.  A rough estimate of the speed of shear
waves (at least in the low shear rate limit) can be obtained from
$\sqrt{G/\rho}$, where $G$ is the static shear modulus and $\rho$ is
the mass density.  A more precise way to measure $u_s$ is to calculate
the transverse current correlation function $C_T(k,\nu)$
\cite{remark4} as a function of frequency $\nu$ and wave number
$k=2\pi n / L_x$ ($n=$ integer) and determine the slope of the
resulting dispersion relation $\nu(k)$ \cite{hansen}.  One of the
novel aspects of our previous work was that we showed that underdamped
systems would be extremely susceptible to nonlinear velocity profiles
near random close packing since the critical velocity $u_c$ tends to
zero at $\phi_{\rm rcp}$.  We therefore predicted that any $u>0$ would
give rise to nonlinear velocity profiles in these systems near
$\phi_{\rm rcp}$.

In these prior studies, we measured the velocity, packing fraction,
and mean-square velocity fluctuation profiles after shearing the
system for a strain of at least $5$ and times $t > t_s$ so that the
shear stress had relaxed to its long-time average value.  This
protocol for bringing sheared systems to steady-state is typical in
both simulations and experiments.  If $t_s$ were the only relevant
time scale, the nonlinear velocity profiles that occur in
boundary-driven planar shear flow when $u>u_c$ would be stable over
long times.  However, in more recent studies, we have found that these
nonlinear velocity profiles are not stable at long times $t \gg t_s$
and slowly evolve toward linear profiles.  In
Fig.~\ref{fig:vp_strain}, we show the slow time evolution of the
velocity profile in a system sheared at $u=0.364$, $\phi=0.90$, and
$e=0.92$.  Note that the strains beyond which the profiles become
linear are extremely large, $\gamma > 25$.  We will also show below
that the time required for the velocity profile to become linear
increases as the system becomes more elastic.  Most of our previous
work focused on nearly elastic systems with $e \approx 0.98$ and
timescales near $t_s$, which made it difficult to detect the profile's
slow evolution.  It would be interesting to know whether this slow
evolution can be seen in experiments on sheared granular media, or
other particulate systems.  However, few experiments have studied such
large strain and time scales.  We note that recent experiments
\cite{mueggenburg} on granular media undergoing planar cyclic shear do
show slow evolution; however, further experiments are required.  Similarly,
simulations of frictional granular media undergoing Couette flow in 3D
have reported significant time evolution of the measured velocity
profiles \cite{baran}.

To quantify the shape of the velocity profiles as a function of time,
we define the degree of nonlinearity as
\begin{equation}
\Delta(t) = \left| 1 - \frac{\langle v_x(y,t) \rangle}
{\langle v_x(y) \rangle_t} \right|,
\end{equation}
where $\langle v_x(y) \rangle_t$ is the time and ensemble average of
the velocity in the flow direction after the evolution of the profile
has ended.  To measure the degree of nonlinearity, we calculated
$\langle \Delta(t) \rangle_y$ averaged over the central part of the
system excluding layers that are immediately adjacent to the top and
bottom walls.  $\langle \Delta(t) \rangle_y \approx 0$ corresponds to
a nearly linear velocity profile, while $\langle \Delta(t) \rangle_y
\approx 1$ is highly nonlinear.

\begin{figure}
\scalebox{0.5}{\includegraphics{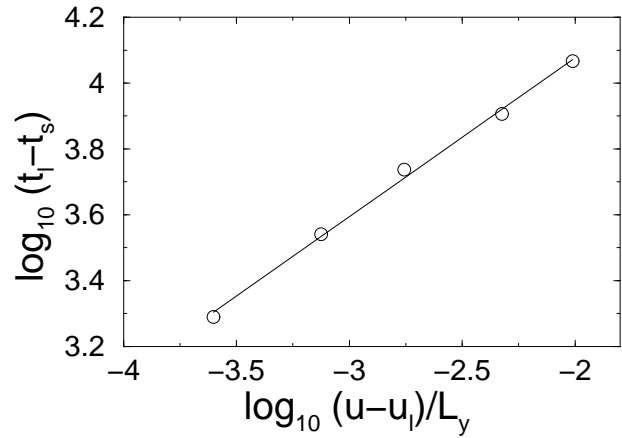}}%
\vspace{-0.15in}
\caption{\label{fig:tl_u} Time $t_l$ required for the velocity
profiles to reach an approximately linear profile relative to the time
$t_s$ required for a shear wave to traverse the system, as a function
of the velocity of the shearing boundary $u-u_l$.  $u_l$ is the
velocity of the shearing boundary below which $t_l = t_s$.  The solid
line has slope $0.5$.  The system parameters are $\phi=0.90$ and $b_n
= 0.0375$.}
\vspace{-0.1in} 
\end{figure}
     
In Fig.~\ref{fig:vevol}, we show $\langle \Delta(t) \rangle_y$ for a
system with $\phi=0.90$ and $e=0.92$ at several different velocities
of the shearing boundary.  At each $u$, the velocity profiles slowly
evolve toward linear profiles and $\langle \Delta(t) \rangle_y$ decays
to zero at long times.  To characterize the long-time behavior, we
define a timescale $t_l$ as the time required for $\langle \Delta(t)
\rangle_y$ to decay to $0.05$, which is slightly above the noise
level.  By definition, the velocity profiles are steady and linear
for times $t > t_l$.  At large $u$, the velocity profiles approach the
linear profile from below as shown in Fig.~\ref{fig:vp_strain}.  At
small $u$, $\langle \Delta(t) \rangle_y$ decays to zero quickly, but
the profiles jump above and below the linear profile as $t \rightarrow
t_l$.  The fact that the velocity profile has significant excursions
above and below linearity at short times signals that the system is in
the quasistatic flow regime.  This behavior at small $u$ is
interesting but not the topic of this study.

\begin{figure}
\scalebox{0.4}{\includegraphics{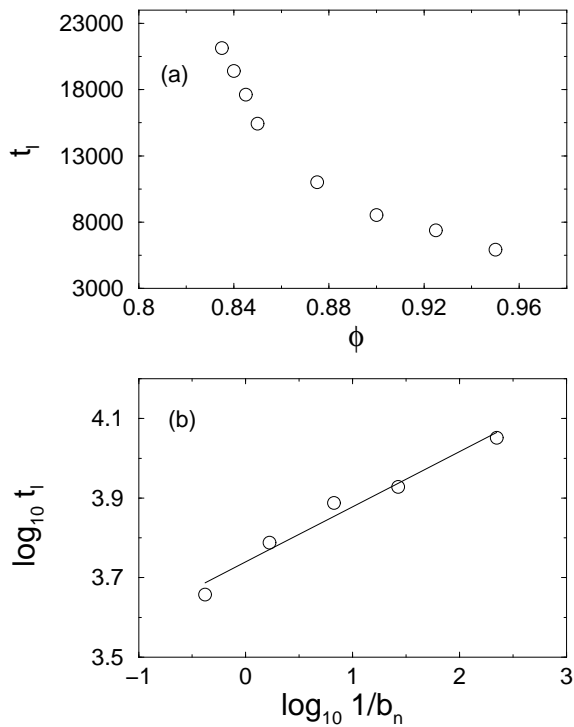}}%
\vspace{-0.1in}
\caption{\label{fig:tl} Time $t_l$ required for the velocity profiles
to reach an approximately linear profile versus (a) $\phi$ at fixed
$b_n=0.0375$ and (b) inverse damping coefficient $1/b_n$ at fixed
$\phi=0.90$ for systems sheared at $u=0.364$.  In panel (b), the solid
line has slope $0.14$.}
\vspace{-0.1in} 
\end{figure}

In Fig.~\ref{fig:tl_u}, we show that the time $t_l$ required for the
velocity profile to become linear increases with the velocity $u$ of
the shearing boundary.  More precisely, $t_l-t_s$ appears to scale as a 
power-law in $u-u_l$
\begin{equation}
\label{tl}
t_l - t_s \sim \left( u-u_l \right)^\beta,
\end{equation} 
for $u> u_l$, where $\beta \approx 0.48 \pm 0.02$.  For $u<u_l$, we
find that the velocity profiles are nonlinear only for times $t < t_s$,
and are linear for all subsequent times.

We also investigated the sensitivity of $t_l$ to changes in the
packing fraction and damping coefficient.  In Fig.~\ref{fig:tl} (a),
we show that $t_l$ increases sharply near random close packing
$\phi_{\rm rcp}$ at small damping.  This rise in $t_l$ is suppressed
at large damping coefficients as shown in Fig.~\ref{fig:tl}.  The
rapid rise in $t_l$ at least for underdamped systems at densities
below random close packing again suggests that these systems are
susceptible to the formation of strongly nonlinear velocity profiles.

These results are surprising and raise an important question regarding
the physical mechanism that is responsible for the slow evolution of
the velocity profiles.  As a first step in addressing this question,
we show below that granular temperature differences across the system
give rise to nonlinear velocity profiles, and that, if a sufficiently
large granular temperature difference can be maintained, nonlinear
velocity profiles will be stable at long times.

\begin{figure}
\scalebox{0.5}{\includegraphics{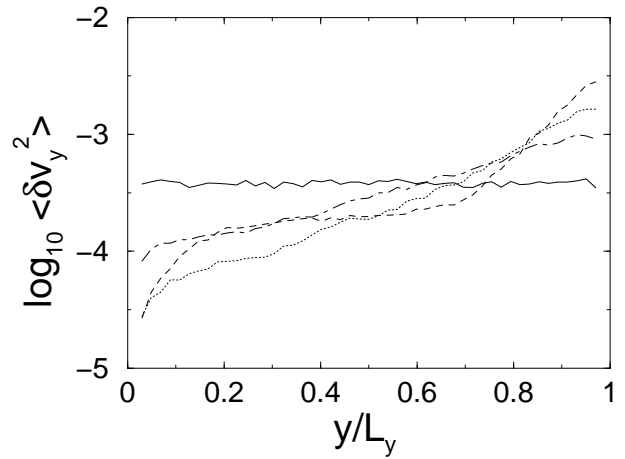}}%
\vspace{-0.15in}
\caption{\label{fig:KEy_strain} Ensemble averaged mean-square velocity
fluctuations $\langle \delta v_y^2 \rangle$ in the shear gradient
direction versus height $y/L_y$ measured from the stationary boundary for the
same system shown in Fig.~\ref{fig:vp_strain}.  The times shown are
$t=520$ (dashed line), $1000$ (dotted line), $2000$ (dot-dashed line),
and $20000$ (solid).  The mean-square velocity fluctuations become
uniform at long times.}
\vspace{-0.1in} 
\end{figure}

\subsection{Combining vibration and shear: Stabilizing nonlinear velocity 
profiles at long times}
\label{sec:vibration}

Recent experiments on sheared granular materials have shown that
strongly nonlinear velocity profiles are accompanied by spatially
dependent granular temperature profiles \cite{losert}.  Thus, an
important question to ask is what role does the granular temperature
play in determining the shape of the velocity profiles in sheared
granular systems.  Also, do these systems require a sufficiently large
granular temperature difference across the system to possess strongly
nonlinear velocity profiles?  Our results in Fig.~\ref{fig:KEy_strain}
suggest that the shapes of the granular temperature and velocity
profiles are strongly linked.  In this figure, we show the time
evolution of the mean-square velocity fluctuations in the
shear-gradient direction, $\langle \delta v_y^2 \rangle$, following the
initiation of shear.  The sequence of times is identical to that shown
in Fig.~\ref{fig:vp_strain}.  At short times, there is a large
difference in the mean-square velocity fluctuations between the `hot'
shearing boundary and `cold' stationary boundary, and the velocity
profile is highly nonlinear.  In contrast, at long times, the
mean-square velocity fluctuations are uniform and the velocity profile
is linear.  Note that in the systems studied here, the granular
temperature difference across the system, $\Delta T$, is roughly equal
to the granular temperature near top boundary, since the mean-square
velocity fluctuations are much smaller near the bottom stationary
boundary.

Figs.~\ref{fig:vp_strain} and~\ref{fig:KEy_strain} show that large
granular temperature differences and nonlinear velocity profiles occur
together.  However, under steady shear, the granular temperature
profiles become uniform and the velocity profiles become linear at
long times.  To further investigate the connection between the
granular temperature and velocity profiles, we study systems that are
both sheared and vibrated, and are thus designed so that granular
temperature differences across the system can be maintained at long
times.  As discussed above in Sec.~\ref{sec:methods}, in our second
set of simulations we drive the top boundary at constant horizontal
velocity $u$, while the location of each particle in the top or bottom
boundary oscillates sinusoidally in time with fixed amplitude $A$ and
frequency $\omega$.  The vertical vibrations cause the mean-square
velocity fluctuations and packing fraction to become spatially
nonuniform with higher velocity fluctuations and lower packing
fraction or dilatancy near the vibrated boundary.

\begin{figure}
\scalebox{0.5}{\includegraphics{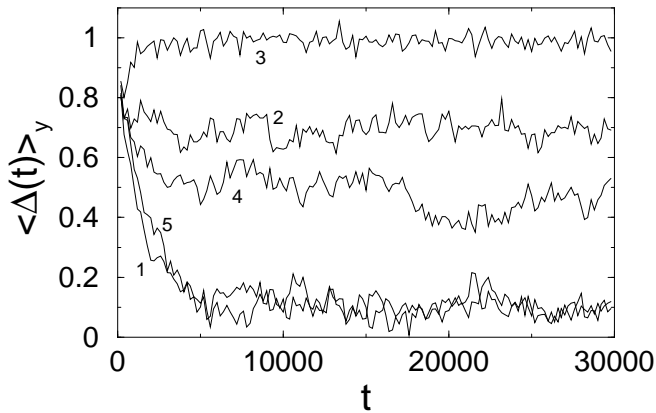}}%
\vspace{-0.15in}
\caption{\label{fig:tc_freq.g0.001} The degree of nonlinearity
$\langle \Delta(t) \rangle_y$ as a function of time for a system that
is both sheared and vibrated with $\phi=0.90$, $u=0.0727$, $e=0.92$,
and $\omega = 0.1$, $0.4$, $0.8$, $1.4$, and $1.8$.  The indexes $1$
to $5$ label the frequencies from smallest to largest.}
\vspace{-0.1in} 
\end{figure}

Fig.~\ref{fig:tc_freq.g0.001} clearly demonstrates that vertical
vibration coupled with shear flow gives rise to {\it stable} nonlinear
velocity profiles at long times.  At low and also at high vibration
frequencies, the degree of nonlinearity $\langle \Delta(t) \rangle_y$
of the velocity profile decays to small values at long times, as we
found previously in Fig.~\ref{fig:vevol} for the unvibrated systems.
In contrast, at frequencies near $\omega_c$, $\langle \Delta(t)
\rangle_y$ is nonzero at long times.  We note that the longest times
shown in Fig.~\ref{fig:tc_freq.g0.001} are at least $10$ times longer
than $t_l$ in the corresponding unvibrated system, confirming
that these are indeed steady-state results.

\begin{figure}
\scalebox{0.5}{\includegraphics{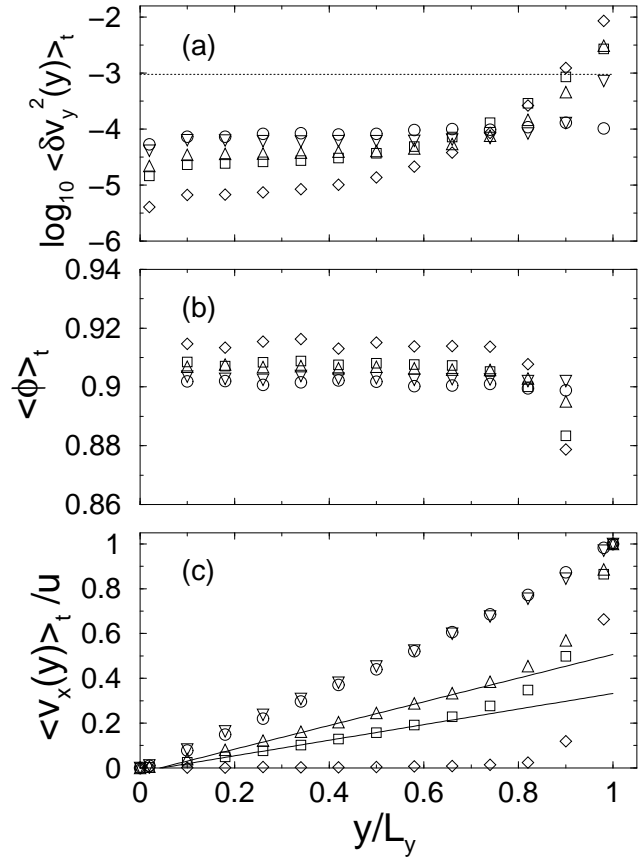}}%
\vspace{-0.15in}
\caption{\label{fig:KEy_freq.g0.001} Time and ensemble averaged
profiles for (a) velocity fluctuations $\langle \delta v_y^2
\rangle_t$, (b) packing fraction $\langle \phi \rangle_t$, and (c)
flow velocity $\langle v_x \rangle_t/u$ for a system at $\phi = 0.90$
and $b_n = 0.0375$ with the top boundary moving at $u=0.0727$ and
vibrated at $\omega=0.1$ (circles), $0.4$ (squares), $0.8$ (diamonds),
$1.4$ (upward triangles), and $1.8$ (downward triangles).  The dotted
line in panel (a) shows the value of the glass transition temperature
in a quiescent system at the same average packing fraction.  The solid
lines in panel (c) show numerical fits to the linear portions of the
velocity profiles at $\omega = 0.4$ and $1.4$.}
\vspace{-0.3in} 
\end{figure}

The results presented in Figs.~\ref{fig:KEy_freq.g0.001} (a) and (b)
suggest that large and sustained granular temperature differences
across the system and the resulting dilatancy are responsible for
stable nonlinear velocity profiles.  This figure shows that nonlinear
velocity profiles occur when there are large granular temperature
differences at $\omega = 0.4$, $0.8$, and $1.4$, while linear profiles
are found when the velocity fluctuations are uniform at $\omega =
0.1$.  Moreover, the degree of nonlinearity in the velocity profiles
increases with the magnitude of the granular temperature difference
across the system.  Fig.~\ref{fig:KEy_freq.g0.001} (a) also shows
the glass transition temperature at the same average packing fraction; we
discuss the relevance of this temperature in more detail in
Sec.~\ref{sec:mechanism}~\cite{remark}.

The granular temperature difference across the system increases with
$\omega$ for $\omega < \omega^*$ but decreases when $\omega >
\omega^*$ with $\omega^* \approx \omega_c$.  The largest granular
temperature difference occurs near the natural frequency $\omega_c$ of
the repulsive spring interactions \cite{remark2}.  The decrease for
$\omega > \omega^*$ occurs because particles adjacent to the vibrating
boundary do not have enough time to react to the collision with the
boundary before another collision occurs.  As a result, vibrations at
large frequencies simply reduce the effective height of the system by
the amplitude of the vibration but do not induce large granular
temperature differences.  We note that $\omega^*$ does not appear to
depend on the shearing velocity $u$.

It is important to emphasize the fact that differences in the
mean-square velocity fluctuations across the system, not the magnitude
of the fluctuations themselves, are important in determining the shape
of the velocity profiles.  For example, a system with the same
interactions, average density, and {\it uniform} temperature $\langle
\delta v_y^2 \rangle > 10^{-3}$ will possess a linear velocity profile
when sheared over the same range of $u$.  Similarly, the system vibrated at
$\omega = 0.1$ (shown as circles in Fig.~\ref{fig:KEy_freq.g0.001}) is
in a glassy state with small but relatively uniform mean-square velocity
fluctuations $\langle \delta v_y^2 \rangle \approx 10^{-4}$ and also
possesses a linear velocity profile.

\begin{figure}
\scalebox{0.5}{\includegraphics{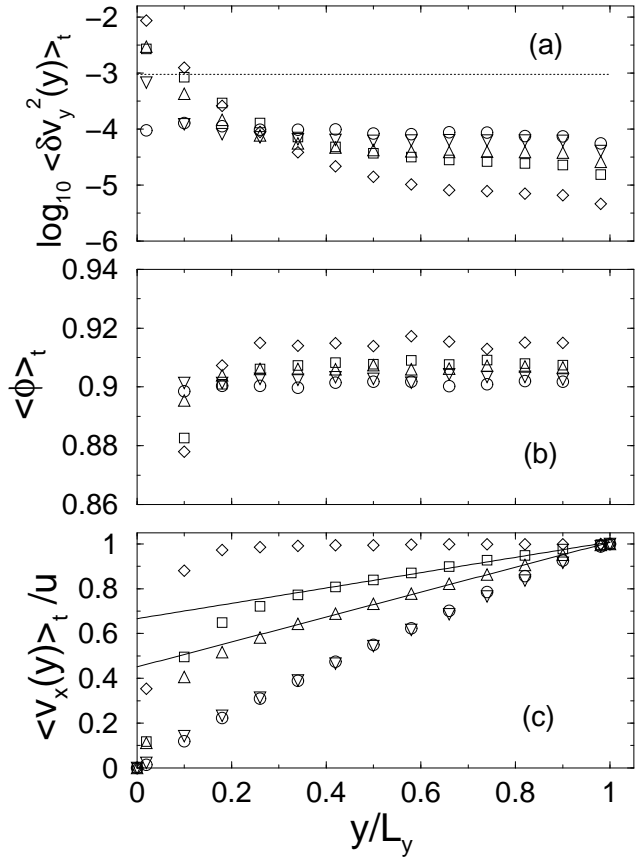}}%
\vspace{-0.15in}
\caption{\label{fig:KEy_freq_bottom.g0.001} Time and ensemble averaged
profiles for (a) velocity fluctuations $\langle \delta v_y^2 \rangle_t$
in the shear gradient direction, (b) packing fraction $\langle \phi
\rangle_t$, and (c) velocity $\langle v_x \rangle_t/u$ in the shear flow
direction for the same system in Fig.~\ref{fig:KEy_freq.g0.001} except
the bottom (not the top) boundary vibrates vertically at $\omega=0.1$
(circles), $0.4$ (squares), $0.8$ (diamonds), $1.4$ (upward
triangles), and $1.8$ (downward triangles). The solid lines in panel
(c) show numerical fits to the linear portions of the velocity
profiles at $\omega = 0.4$ and $1.4$.}
\vspace{-0.1in} 
\end{figure}

In systems that possess nonlinear velocity profiles, we find that the
largest local shear rate does not occur equally likely at both
boundaries.  Instead, the portion of the system with the largest local
shear rate always forms near the boundary with the largest $\langle
\delta v_y^2 \rangle$ and resulting dilatancy.  This is confirmed in
Fig.~\ref{fig:KEy_freq_bottom.g0.001}, which shows results in a system
that is identical to that in Fig.~\ref{fig:KEy_freq.g0.001} except the
bottom wall, not the top wall, is vibrated.  The vertical vibrations
induce a larger granular temperature near the bottom vibrated but
unsheared boundary.  Thus, the largest local shear rates occur near
the bottom boundary, as shown for $\omega=0.4$, $0.8$, and $1.4$ in
Fig.~\ref{fig:KEy_freq_bottom.g0.001} (c).

\section{Discussion}
\label{sec:mechanism}

In this section, we will focus on several aspects of our results in
more detail.  First, we find that nonlinear velocity profiles form
only when the difference in the granular temperature across the system
exceeds a threshold value $\Delta T > \Delta T_0$.  Second, in systems
with large granular temperature differences $\Delta T > \Delta T_0$ at
long times, the velocity profiles are linear near the `cold' wall but
highly nonlinear near the `hot' wall.  These profiles differ in shape
from those found in the sheared but unvibrated systems \cite{xu}.
Finally, we point out that shear bands form when the average shear
stress of the system $\langle \sigma_{xy} \rangle_t$ falls below the
yield stress $\sigma_0$ required to initiate sustained flow in 
a static system.

Figs.~\ref{fig:KEy_freq.g0.001} and~\ref{fig:KEy_freq_bottom.g0.001}
clearly show that differences in the mean-square velocity fluctuations
across the system give rise to nonlinear velocity profiles.  However,
our results indicate that the granular temperature difference required
to generate a nonlinear velocity profile must exceed a threshold
value.  For example, systems with vibration frequency $\omega=1.8$
(downward triangles) in Figs.~\ref{fig:KEy_freq.g0.001}
and~\ref{fig:KEy_freq_bottom.g0.001} have relatively large granular 
temperature differences (slightly less than $\Delta T = 10^{-3}$), but
possess nearly linear velocity profiles.  In contrast, systems with
$\Delta T > 10^{-3}$ (for example $\omega = 0.4$, $0.8$, and
$1.4$) possess strongly nonlinear velocity profiles.

The threshold granular temperature difference, which is roughly equal
to the granular temperature near the vibrated boundary, appears to agree
with the glass transition temperature for a quiescent equilibrium
system at the same average packing fraction \cite{remark}.  This
correspondence seems reasonable since in equilibrium systems the
temperature must exceed the glass transition temperature to cause
large density fluctuations.  Similarly, in sheared dissipative
systems, it is difficult to create sufficiently large density
gradients required for nonlinear velocity profiles if the granular
temperature difference is below a threshold $\Delta T_0$.

We also find that the nonlinear velocity profiles that occur in the
sheared and vibrated athermal systems have qualitatively different
shapes compared to those found for the sheared but unvibrated systems.  For
example, the nonlinear velocity profiles for $\omega = 0.4$ (squares)
and $1.4$ (upward triangles) in Fig.~\ref{fig:KEy_freq.g0.001} are
composed of a linear portion that extends from the bottom boundary at
$y=0$ to $y \sim 0.8$, and a strongly nonlinear part near the top
boundary.  We note that in vibrated systems the crossover between
the linear and highly sheared behavior occurs where the local packing
fraction switches from uniform to nonuniform.

Finally, we will address an interesting claim made in
Ref.~\cite{varnik} that shear bands form in sheared glassy systems
when the average shear stress in the system falls below the yield
shear stress required to induce flow in a static state.  In systems
that form shear bands, shear flow is confined to a small portion of
the system while the remainder of the system remains nearly static.
In contrast, all parts of the system flow when systems possess generic
nonlinear velocity profiles.

\begin{figure}
\scalebox{0.5}{\includegraphics{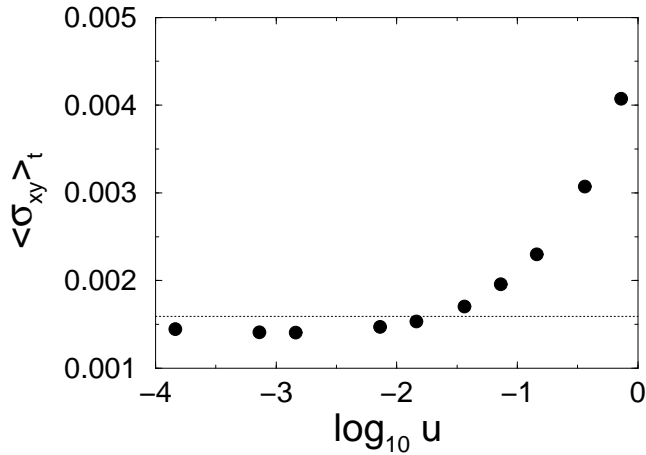}}%
\vspace{-0.15in}
\caption{\label{fig:stress_u} Time averaged shear stress $\langle
\sigma_{xy} \rangle_t$ versus the boundary velocity $u$ for a sheared but
unvibrated system at $\phi=0.90$ and $b_n=0.0375$.  The dotted line
indicates the yield stress $\sigma_0$ required to initiate sustained
flow in a static system at the same packing fraction.}
\vspace{-0.1in} 
\end{figure}

Does the constraint on the average shear stress guarantee that shear
bands will occur in the repulsive athermal systems studied here?  In
Fig.~\ref{fig:stress_u}, we compare the time-averaged shear stress
$\langle \sigma_{xy} \rangle_t$ sheared at fixed $u$ to the yield
shear stress $\sigma_0$ required to induce flow in an originally
static unsheared state at the same $\phi$.  We find that at small
boundary velocities $u$, $\langle \sigma_{xy} \rangle_t$ is slightly
below $\sigma_0$.  However, even though the average shear stress is
below the yield stress, the velocity profiles are linear for long
times (as shown in Fig.~\ref{fig:vevol}) and not highly localized
or shear-banded.

Thus, the condition $\langle \sigma_{xy} \rangle_t < \sigma_0$ alone
does not ensure that sheared repulsive athermal systems will form
shear bands.  An additional requirement must be satisfied to stabilize
shear bands at long times---the systems must possess sufficiently
large granular temperature differences.  It should be noted, however,
that the differences between the average and the yield shear stress
are small in the systems we studied and that shear stress fluctuations
may inhibit the formation of shear bands.  Thus, more work should be
performed to verify the presented results.  In fact, we are now
attempting to determine the variables that set the difference between
$\langle \sigma_{xy} \rangle_t$ and $\sigma_0$ and whether this
difference persists in the large system limit.

\begin{figure}
\scalebox{0.5}{\includegraphics{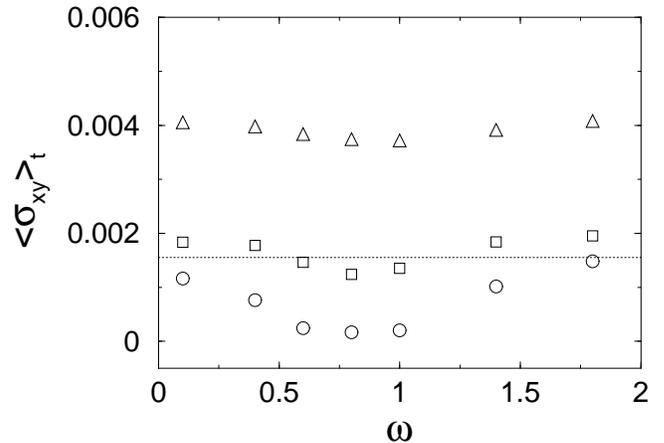}}%
\vspace{-0.15in}
\caption{\label{fig:stress_freq} Time averaged shear stress $\langle
\sigma_{xy} \rangle_t$ for the sheared and vibrated system at
$\phi=0.90$ and $b_n=0.0375$ as a function of $\omega$ for $u=0.00727$
(circles), $0.0727$ (squares), and $0.727$ (upward triangles).  The
dotted line indicates the yield shear stress $\sigma_0$ for the static
system at the same average density and dissipation.}
\vspace{-0.1in} 
\end{figure}

\begin{figure}
\scalebox{0.5}{\includegraphics{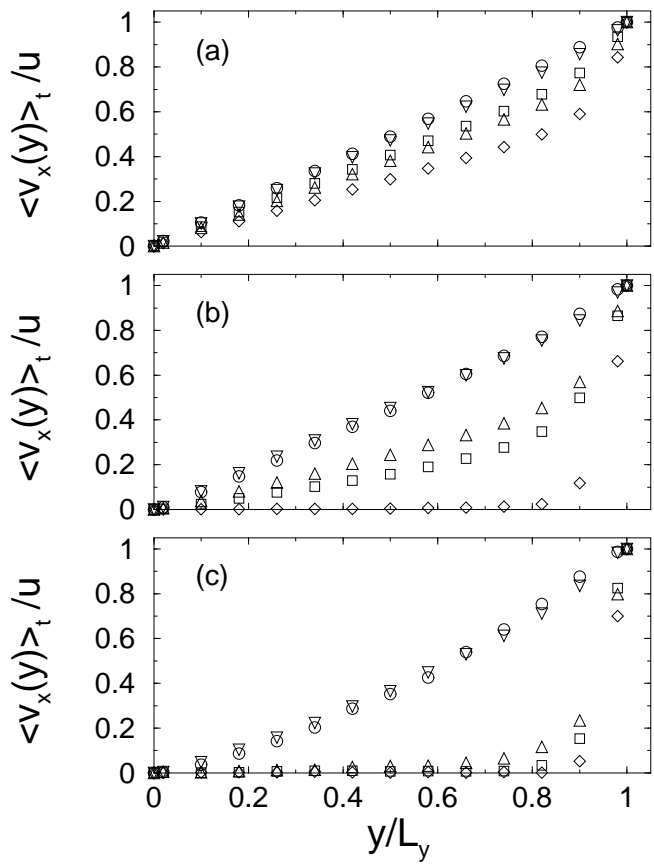}}%
\vspace{-0.15in}
\caption{\label{fig:shear_band} Time averaged velocity profiles
$\langle v_x(y) \rangle_t$ in the sheared and vibrated systems at
$\phi = 0.90$ and $b_n = 0.0375$ for several vibration frequencies
$\omega=0.1$ (circles), $0.4$ (squares), $0.8$ (diamonds), $1.4$
(upward triangles), and $1.8$ (downward triangles) at (a) $u=0.727$,
(b) $0.0727$, and (c) $0.00727$.}
\vspace{-0.1in} 
\end{figure}

Fig.~\ref{fig:stress_freq} shows that the average shear stress falls
below the yield shear stress in the sheared and vibrated systems over
a range of frequencies at $u=0.00727$ and $0.0727$, but not at
$0.727$.  We have also confirmed that the granular temperature
differences in these systems satisfy $\Delta T > \Delta T_0$ over the
range of frequencies $0.4 < \omega < 1.4$. Therefore, from the
discussion above, one expects shear bands to form for the two higher
shear rates, but not the lowest one. This prediction is confirmed in
Fig.~\ref{fig:shear_band}.  At $u=0.727$, nonlinear velocity profiles
form, but shear is not highly localized into a shear band.  In
contrast, at $u=0.00727$ there is a range of frequencies $0.4 < \omega
< 1.4$ over which shear bands form.

\section{Conclusions}
\label{sec:conclusions}

In this article we reported on recent simulations of model frictionless
granular systems undergoing boundary-driven planar shear flow in 2D over a
range of flow velocities and average densities, and for particles with
varying degrees of inelasticity.  These studies have produced several
interesting and novel results that are relevant to a variety of jammed
and glassy systems subjected to planar shear flow.  First, we find that
nonlinear velocity profiles are not stable at long times.
Nonlinear velocity profiles form when the boundary velocity
exceeds a characteristic speed set by the shear wave speed in the
material, but they slowly evolve toward linear profiles at long times.
In addition, the granular temperature and packing fraction profiles
are initially spatially dependent but become uniform at long times.
We measured the time $t_l$ required for the velocity profiles to
become linear and for the granular temperature and density to become
homogeneous throughout the system.  We find that $t_l$ increases as
$u^{0.5}$ at large $u$ and increases strongly as $\phi$ approaches
$\phi_{\rm rcp}$ for nearly elastic systems.

These results imply that sufficiently large and sustained granular
temperature differences between the `hot' and `cold' boundaries are
required to stabilize nonlinear velocity profiles at
long times.  We also studied systems in which vertical vibrations of
the top or bottom boundary were superimposed onto planar shear flow to
maintain granular temperature differences across the system.  In the
sheared {\it and} vibrated systems, we find that nonlinear velocity
profiles are stable when the granular temperature difference exceeds a
threshold value $\Delta T_0$, which roughly corresponds to the glass
transition temperature in an equilibrium system at the same average
density.  The nonlinear velocity profiles, however, differ in shape
from those found previously for planar shear flow.  The velocity
profiles are composed of a linear part that exists where the packing
fraction is spatially uniform and a nonlinear portion that exists
where the packing fraction varies strongly in space.  Finally, we have
shown that the nonlinear velocity profiles become highly localized
when the average shear stress in the system is below the yield shear
stress.

\begin{figure}
\scalebox{0.62}{\includegraphics{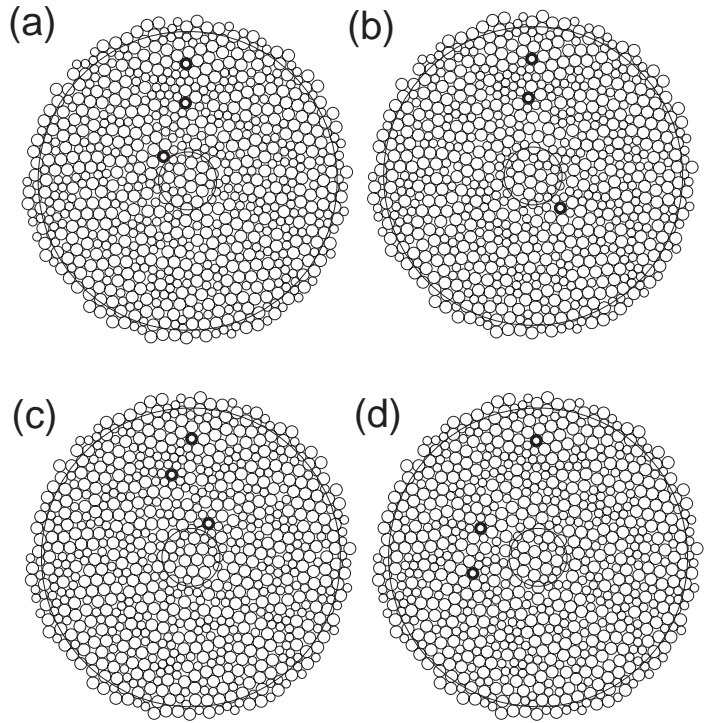}}%
\vspace{-0in}
\caption{\label{fig:frame} Snapshots from 2D systems undergoing
Couette shear flow at $\phi = 0.845$ and $b_n = 0.0375$.
Panels (b), (c), and (d) differ from panel (a) by $2$, $6$, and $19$
rotations, respectively.  All particles with centers inside the inner
ring at $r=R_1$ move at fixed rotation rate $\Omega = 0.01$ counter
clockwise.  All particles with centers outside the outer ring at
$r=R_2$ are stationary.}
\vspace{-0.2in} 
\end{figure}

\section{Future Directions}     
\label{sec:future}
 
Several additional studies are necessary to fully understand
dense shear flows in granular systems, and these will be presented in
future work \cite{xu2}.  First, we intend to investigate the influence
of dynamic and static friction forces on the long-time stability of
nonlinear velocity profiles.  Preliminary results suggest that dynamic
friction is not sufficient to stabilize nonlinear velocity profiles in
systems undergoing planar shear flow.  In initial studies with
weak dissipation and dynamic friction \cite{remark3} near
$\phi_{\rm rcp}$, we found that $t_l$ is similar to that for systems with
dissipation and no dynamic friction.  However, more extensive studies
of dynamic as well as static friction are required to make a
definitive statement about the influence of friction on the long-time
stability of nonlinear velocity profiles \cite{aharonov,volfson}.

\begin{figure}
\scalebox{0.5}{\includegraphics{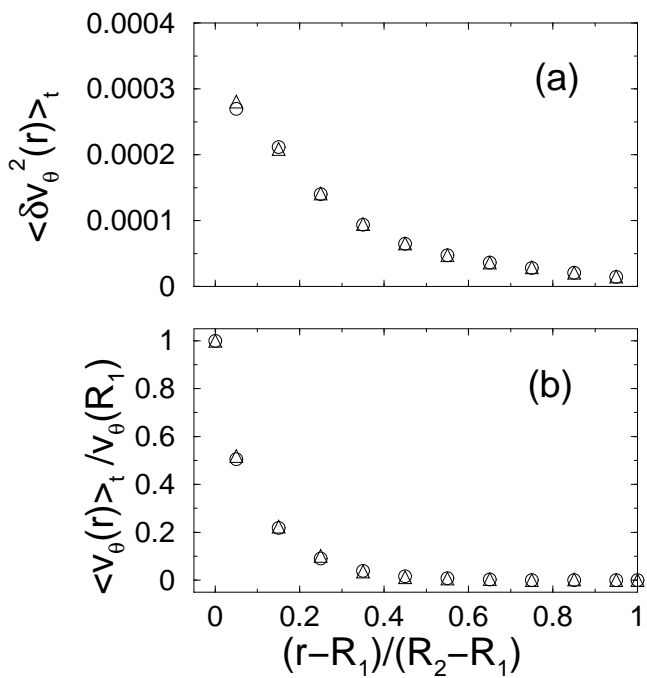}}%
\vspace{-0.in}
\caption{\label{fig:v_T_couette} (a) Mean-square velocity fluctuations
$\langle \delta v_{\theta}^2 (r)\rangle$ in the tangential direction
and (b) tangential velocity $v_{\theta}(r)$ normalized by the speed
$v_{\theta}(R_1)$ at the inner shearing boundary versus the distance
from the inner boundary $(r-R_1)/(R_2 - R_1)$ in a 2D Couette cell
with $\phi =0.845$ and $b_n=0.0375$.  The symbols indicate that time
averages taken between the $20$th and $80$th (circles) or the $60$th
and $80$th (triangles) rotations are nearly identical.}
\vspace{-0.1in} 
\end{figure}

Second, most experimental studies of velocity profiles in granular
systems are performed in (angular) Couette, not planar shear cells.
How does the geometry of the shear cell influence the velocity
profiles?  Are shear bands stable in frictionless, athermal systems
undergoing Couette shear flow?  Fig.~\ref{fig:frame} shows preliminary
results from studies of 2D Couette shear flow in systems at $\phi =
0.845$, $b_n = 0.0375$, and rotation rate $\Omega = 0.01$ in the
counter clockwise direction.  Panels (b), (c), and (d) show snapshots
of the system $2$, $6$, and $19$ rotations after the initial
configuration in panel (a).  The snapshots provided in panels (a) -
(c) reveal that the shear band in this system is approximately $5-8$
small particle diameters wide since the two highlighted particles
closest to the outer boundary do not rotate significantly even after
$6$ rotations of the inner boundary.  A comparison of panels (c) and
(d) shows that at long times the particles in the flowing region are
able to diffuse perpendicular to the boundaries.  The mean-square
velocity fluctuations $\langle \delta v_{\theta}^2 \rangle$ in the
tangential direction and the tangential velocity $v_{\theta}$
normalized by the speed at the inner boundary are shown in
Figs.~\ref{fig:v_T_couette} (a) and (b) as a function of the distance
from the inner boundary $(r-R_1)/(R_2 - R_1)$.  Averages over varied
numbers of rotations demonstrate that the nonlinear velocity profiles
are stable at long times.  As we found in our studies of planar shear
flow, nonlinear velocity profiles occur when the granular temperature
is nonuniform.  However, in Couette shear flows, the shear stress is
also spatially dependent.  Thus, the distinct contributions from
nonuniform shear stress and nonuniform granular temperature need to be
disentangled.  In future work, we will determine whether nonlinear
velocity profiles persist when we vibrate the outer boundary to create
a more uniform granular temperature profile.
  
Finally, there have been several recent computational studies of
effective temperatures defined from fluctuation dissipation relations,
linear response theory, and elastic energy fluctuations in dense
granular systems \cite{makse,kondic,ohernt}.  These studies have shown
that a consistent effective temperature can be defined for dense shear
flows, i.e. the effective temperatures $T_{\rm eff}$ obtained from the
above definitions agree with each other but $T_{\rm eff}$ is much
larger than the granular temperature of the system.  Because $T_{\rm
eff}$ describes fluctuations on long length and time scales, it is
possible that the effective temperature will play a significant role
in determining the shape of velocity profiles.  Thus, it is important
to study the effective temperature as well as the granular temperature
in sheared glassy and athermal systems.

\subparagraph{Acknowledgments}

We thank J. Blawzdziewicz, A. Liu, and S. Nagel for helpful comments.
Financial support from NSF DMR-0448838 (NX,CSO), NASA NNC04GA98G (LK),
and the Kavli Institute for Theoretical Physics under NSF PHY99-07949
(CSO,LK) is gratefully acknowledged.  We also thank Yale's High
Performance Computing Center for generous amounts of computing time.


\begin{references}

\bibitem{howell} D. Howell, R. P. Behringer, and C. Veje, {\it
Phys. Rev. Lett.} {\bf 82}, 5241 (1999); C. T. Veje, D. W. Howell, and
R. P. Behringer, {\it Phys. Rev. E} {\bf 59}, 739 (1999).

\bibitem{losert}
W. Losert, L. Bocquet, T. C. Lubensky, and J. P. Gollub,
{\it Phys. Rev. Lett.} {\bf 85}, 1428 (2000). 

\bibitem{mueth2}
D. M. Mueth, {\it Phys. Rev. E} {\bf 67}, 011304 (2003).

\bibitem{mueth} 
D. M. Mueth, G. F. Debregeas, G. S. Karczmar,
P. J. Eng, S. R. Nagel, and H. M. Jaeger, {\it Nature} {\bf 406}, 385
(2000). 

\bibitem{mueggenburg}
N. W. Mueggenburg, {\it Phys. Rev. E} {\bf 71}, 031301 (2005).

\bibitem{fenistein}
D. Fenistein and M. van Hecke, {\it Nature} {\bf 425}, 256 (2003).

\bibitem{pouliquen}
O. Pouliquen and R. Gutfraind, {\it Phys. Rev. E}, 552 (1996).

\bibitem{jenkins}
J. T. Jenkins, ``Rapid granular shear flows driven by identical, bumpy
frictionless boundaries,'' in {\it Powders and Grains 93}, ed. C. Thornton
(Balkema, Rotterdam, 1993).     

\bibitem{wildman}
R. D. Wildman, personal communication.

\bibitem{thompson}
P. A. Thompson and G. S. Grest, {\it Phys. Rev. Lett.} {\bf 67}, 1751 (1991).

\bibitem{aharonov}
E. Aharonov and D. Sparks, {\it Phys. Rev. E} {\bf 65}, 051302 (2002).

\bibitem{volfson}
D. Volfson, L. S. Tsimring, and I. S. Aranson, {\it Phys. Rev. E}
{\bf 68}, 021301 (2003); {\it Phys. Rev. E} {\bf 69}, 031302 (2004).

\bibitem{dacruz} F. da Cruz, S. Emam, M. Prochnow, J.-N. Roux, and
F. Chevoir, cond-mat/0503682, (2005).

\bibitem{xu}
N. Xu, C. S. O'Hern, and L. Kondic,
{\it Phys. Rev. Lett.} {\bf 94}, 016001 (2005).

\bibitem{ohern}
C. S. O'Hern, S. A. Langer, A. J. Liu, and S. R. Nagel,
{\it Phys. Rev. Lett.} {\bf 88}, 075507 (2002); C. S. O'Hern, L. E. 
Silbert, A. J. Liu, and S. R. Nagel, {\it Phys. Rev. E} {\bf 68}, 011306
(2003).

\bibitem{numrec} 
W. H. Press, B. P. Flannery, S. A. Teukolsky, and W. T. Vetterling, 
{\it Numerical Recipes in Fortran 77} (Cambridge University Press, 
New York, 1986). 

\bibitem{herrmann} H. J. Herrmann and S. Luding, {\it Continuum
Mech. Thermodyn.} {\bf 10}, 189 (1998).

\bibitem{varnik}
F. Varnik, L. Bocquet, J.-L. Barrat, and L. Berthier,
{\it Phys. Rev. Lett.} {\bf 90}, 095702 (2003).

\bibitem{remark4} The transverse current correlation function is
defined by $C_T(k,t) = \frac{k^2}{N} \langle j_{\perp}(k,t)
j_{\perp}(-k,0) \rangle$, where $j_{\perp}(k,t) = \sum_{l=1}^N
v_{l,y}(t) \exp\left[ -i k x_l(t) \right]$.  The dependence on
frequency $\nu$ can be obtained by Fourier transforming $C_T(k,t)$.

\bibitem{hansen}
J. P. Hansen and I. R. McDonald, {\it Theory of Simple Liquids} (Academic 
Press, London, 1986). 

\bibitem{baran}
O. Baran and L. Kondic, {\it Phys. Fluids} (2005).

\bibitem{remark} The glass transition temperature was calculated in
equilibrium systems by measuring the density autocorrelation
$C(k_0,t)$ (at a wavenumber $k_0$ near the peak in the static
structure factor) as a function of decreasing temperature.  We made
small temperature jumps and equilibrated the system at each
temperature.  We measured the structural relaxation time $\tau$---the
time at which $C(k_0,t)$ decayed to $1/e$---as a function of temperature
and identified the glass transition temperature by fitting $\tau(T)$
to a Vogel-Fulcher form \cite{vogel}.

\bibitem{remark2} More precisely, the system is composed of $N/2$
large and $N/2$ small particles with diameter ratio $1.4$, and thus we
expect the largest granular temperature difference across the system
to occur at frequencies near $\omega^* = \omega_c/1.2$.

\bibitem{vogel}
M. D. Ediger, C. A. Angell, and S. R. Nagel, J. Phys. Chem. {\bf 100}, 
13200 (1996)

\bibitem{xu2}
N. Xu, C. S. O'Hern, and L. Kondic, (unpublished).

\bibitem{remark3} Granular systems with both dissipation and dynamic
friction have nonconservative pair forces that are nonzero when two
particles overlap $-b_n \left[ {\vec v}_{ij} \cdot {\hat
r}_{ij}\right] {\hat r}_{ij} - b_t \left[ {\vec v}_{ij} \cdot {\hat
t}_{ij} \right] {\hat t}_{ij}$, where ${\hat t}_{ij} \cdot {\hat
r}_{ij} = 0$.  When $b_t \ne 0$, time evolution of the rotational
degrees of freedom must also be considered.

\bibitem{makse}
H. A. Makse and J. Kurchan, {\it Nature} {\bf 415}, 614 (2002).

\bibitem{kondic}
L. Kondic and R. P. Behringer, {\it Europhys. Lett.} {\bf 67}, 205 (2004).

\bibitem{ohernt}
C. S. O'Hern, A. J. Liu, and S. R. Nagel, {\it Phys. Rev. Lett.} {\bf 93}, 
165702 (2004).

\end{references}
\end{document}